\documentclass[aoas,MSNbibl,nameyear,dvips]{arximspdf}
\usepackage{dcolumn}
\usepackage{graphicx}

\doi{10.1214/13-AOAS632} 
\volume{7}
\issue{3}
\pubyear{2013}
\firstpage{1450}
\lastpage{1473}

\makeatletter
\newcolumntype{d}[1]{D{.}{.}{#1}}
\makeatother

\begin{document}
\begin{frontmatter}

\title{Assessment of mortgage default risk via Bayesian state space models}
\runtitle{Bayesian mortgage default risk}

\begin{aug}
\author[a]{\fnms{Tevfik} \snm{Aktekin}\ead[label=e1]{tevfik.aktekin@unh.edu}},
\author[b]{\fnms{Refik} \snm{Soyer}\corref{}\ead[label=e2]{soyer@gwu.edu}\thanksref{t1}}
\and
\author[c]{\fnms{Feng} \snm{Xu}\ead[label=e3]{fxu@canes.gsw.edu}}
\thankstext{t1}{Supported by the NSF Grant DMS-09-1516 with George Washington University.}
\runauthor{T. Aktekin, R. Soyer and F. Xu}
\affiliation{University of New Hampshire, George Washington University
and Georgia~Southwestern University}
\address[a]{T. Aktekin\\
University of New Hampshire\\
10 Garrison Avenue\\
Durham, New Hampshire 03824\\
USA\\
\printead{e1}}
\address[b]{R. Soyer\\
George Washington University\\
2201 G Street, NW\\
Washington, DC 20052\\
USA\\
\printead{e2}}
\address[c]{F. Xu\\
Georgia Southwestern University\\
800 Georgia Southwestern State University Drive\\
Americus, Georgia 31709\\
USA\\
\printead{e3}}
\end{aug}

\received{\smonth{10} \syear{2012}}
\revised{\smonth{1} \syear{2013}}

%
\begin{abstract}
Managing risk at the aggregate level is crucial for banks and financial
institutions as required by the
Basel III framework. In this paper, we introduce discrete time Bayesian
state space models with Poisson
measurements to model aggregate mortgage default rate. We discuss
parameter updating, filtering,
smoothing, forecasting and estimation using Markov chain Monte Carlo
methods. In addition,
we investigate the dynamic behavior of the default rate and the effects
of macroeconomic variables.
We illustrate the use of the proposed models using actual U.S.
residential mortgage data
and discuss insights gained from Bayesian analysis.
\end{abstract}

%
\begin{keyword}
\kwd{Mortgage default}
\kwd{mortgage risk}
\kwd{Bayesian inference}
\kwd{state space}
\kwd{dynamic Poisson process}
\end{keyword}

\end{frontmatter}

\section{Introduction}\label{sec1}
Given the large size of outstanding residential mortgage loans in the
U.S., a healthy mortgage market is important for stability of the
financial markets and the whole economy. Due to its significant costs
upon mortgage borrowers, lenders, insurers and investors of mortgage
backed securities, management of mortgage default risk is one of the
primary concerns for the policy makers and financial institutions.

Most commonly used measures of mortgage default risk are delinquency
and foreclosure rates of mortgage loans. They provide a general
description of how the mortgage market performs, compared to the macro
economy. According to \citet{GH89}, mortgage default is legally defined
as the transfer of property ownership from the borrower to the lender.
The majority of researchers who focus on modeling of default risk
define mortgage default as being delinquent in a mortgage payment for
90 days as discussed in \citet{AC98}. In this paper we use the latter
definition to distinguish default from foreclosure.

Most of the work in the mortgage default risk literature has focused
mainly on the individual default behavior of borrowers, and the effects
of mortgage loan, property, borrower and economic characteristics on
default risk. \citet{QS92} provide a detailed literature review of
research in mortgage default risk until 1992. More recent developments
can be found in \citet{L04}. There are two dominant classes of models in
the literature. The first class of models is based on the ruthless
default assumption and is option theoretic where the mortgage value,
prepayment and default options are determined via stochastic behavior
of prices and interest rates as in \citet{KKME90}. The second class is
based on the hazard rate models where time to mortgage default is a
random variable with hazard rate as a function of individual borrower
and loan characteristics as studied by Lambrecht, Perraudin and
Satchell (\citeyear{LPS97,LPS03}) and
\citet{SF09}. Both classes of models are based on the behavior of
individual mortgages. But studying the default behavior at the
aggregate level is also of interest to financial institutions and
policy makers to be able to predict default rates and to develop
appropriate mitigation instruments. As pointed out by \citet{T06},
managing risk at the aggregate level is crucial for banks and financial
institutions as required by the Basel III framework which encourages
banks to identify and manage present and future risks. \citet{T06}
models the probability of default at the aggregate level for two
default classes, all-corporate and speculative-grades in U.S. as a
stochastic process.

Modeling aggregate default rates requires consideration of several
issues. First, it is important to identify the effect of macroeconomic
variables on the aggregate default rate. This is pointed out by \citet
{T06} but is not considered in his model. Another issue to assess is if
the aggregate default rate exhibits a dynamic behavior. In modeling
individual default rates, Soyer and Xu (\citeyear{SF09}) point out that default
rates are nonmonotonic. More specifically, the authors report that
default rates are typically first increasing and then decreasing over
the duration of the mortgage. It is not unreasonable to expect that the
aggregate default rate will also follow such a dynamic behavior. Third,
as noted by \citet{K11}, it is not uncommon to have correlated defaults
over time. Thus, it is desirable for models to capture such correlations.

In this paper, we present a discrete time Bayesian state-space model
for Poisson counts to address the above issues. The proposed model
enables us to describe the dynamic behavior of aggregate mortgage
default rates over time and assumes a Markovian structure to describe
the correlated default rates. This Markovian structure enables us to
capture correlations between the number of defaults over time and
provides an alternate way of modeling time-series of counts. Since the
Markovian structure is assumed for the parameter, that is, for the
default rate, our model can be classified as a \textit{parameter driven
Markov model} using the terminology of \citet{Cox81}. We introduce an
extension of the model by modulating the default rate by considering
the effect of covariates describing the economic environment. This
model can be considered a discrete time version of a \textit{modulated
Poisson process model} of \citet{Cox72}. This class of models and their
Bayesian analysis have not been considered in the literature before. To
the best of our knowledge, only a few studies consider Bayesian methods
in modeling mortgage default risk in the literature. \citet{H88}
introduces basic Bayesian concepts and \citet{PPG08} apply Bayesian
methods to forecast mortgage prepayment rates. More recently, \citet
{K10} introduces the incorporation of expert knowledge in estimating
default rates from a Bayesian point of view, details a binomial model
with dependent defaults and discusses implications of such models on
risk management. As noted by \citet{K11}, the Bayesian approach provides
a coherent framework to combine data with prior information and enables
us to make inferences using probabilistic reasoning. As will be
discussed in our illustrations, additional insights are gained from the
Bayesian analysis.

A summary of our paper is as follows: In Section \ref{sec2} we introduce a
Bayesian state space model for the monthly default counts for a given
mortgage pool. Section~\ref{sec3} is dedicated to the development of a discrete
time Bayesian state space model with covariates.
We discuss the Bayesian analysis of the
models in Section \ref{sec4} using Markov chain Monte Carlo methods. An
illustration of the proposed models is presented in Section \ref{sec5} using
real default count data for different mortgage pools where we discuss
both in and out of sample fit issues for our models and compare them
with the Bayesian Poisson regression which we use as a benchmark.
Finally, in Section \ref{sec6} we conclude with a summary of our findings and
suggestions for future work.

\section{A dynamic model for number of defaults}\label{sec2}
We first introduce a discrete time Bayesian model with Poisson
observations and a default rate that evolves over time according to a
Markov process. This model does not take into account the effects of
covariates on the default rate of a given mortgage pool. \citet{SM86}
consider a similar state space model for exponential measurements which
was used by \citet{MS02} in the context of software reliability.

Let $N_{t}$ be the number of defaults of a given mortgage pool during
the month $t$ and $\theta_{t}$ be its default rate for $t=1,\ldots,T$.
Given $\theta_{t}$, we assume that the number of defaults during the
month $t$ is described by a discrete time nonhomogeneous Poisson process,
%
\begin{equation}
(N_{t}|\theta_{t}) \sim \operatorname{Pois}(\theta_{t}).
\label{measurement}
\end{equation}
In (\ref{measurement}) it is assumed that given the default rate $\theta
_{t}$, the default counts ${N_{t}}$s are conditionally independent.
Also, (\ref{measurement}) acts as an observation equation for discrete time.

For the state evolution equation of $\theta_{t}$'s, we assume that
consecutive default rates exhibit a Markovian behavior similar to that
considered by \citet{T06} at the aggregate level. The Markovian
evolution of default rates over time is described by
%
\begin{equation}
\theta_{t}= \frac{\theta_{t-1}}{\gamma} \varepsilon_{t},
\label{state1}
\end{equation}
where $(\varepsilon_{t}|N^{(t-1)}) \sim \operatorname{Beta}[\gamma a_{t-1},(1-\gamma)
a_{t-1}], a_{t-1} > 0, 0 < \gamma< 1$, and\break $N^{(t-1)} = \{
N_{1},\ldots, N_{t-1} \}$. Here, $\gamma$ acts like a discounting
term between consecutive default rates. An evolution structure similar
to (\ref{state1}) was considered by \citet{U97} in the context of
modeling stochastic volatility. A similar setup for a general family of
non-Gaussian models was introduced by \citet{SGF12}. Our model can be
obtained as a special case.

The state equation (\ref{state1}) implies a stochastic ordering between
the default rates, $\theta_{t}<\frac{\theta_{t-1}}{\gamma}$.
Therefore, it can be shown that
%
\begin{equation}
\bigl(\theta_{t}|\theta_{t-1},N^{(t-1)}\bigr) \sim
\operatorname{Beta}\biggl[\gamma a_{t-1}, (1-\gamma ) a_{t-1};
\biggl(0,\frac{\theta_{t-1}}{\gamma}\biggr)\biggr], \label{scaledbeta}
\end{equation}
that is, a truncated Beta density. If one assumes that a priori $\theta
_{0}$ follow a gamma density as
%
\begin{equation}
\bigl(\theta_{0}|N^{(0)}\bigr) \sim \operatorname{Gamma}(a_{0},b_{0}),
\label{prior}
\end{equation}
then one can develop an analytically tractable Bayesian analysis for
the model.
Following \citet{SM86}, as a result of (\ref{state1}) and (\ref{prior})
we can obtain
%
\begin{equation}
\bigl(\theta_{t-1}|N^{(t-1)}\bigr) \sim \operatorname{Gamma}(a_{t-1},b_{t-1}),
\label{prior2}
\end{equation}
which can be shown by induction. Given the measurement equation (\ref
{measurement}), the state evolution equation (\ref{state1}) and the
prior (\ref{prior}), the posterior default rates and one-step-ahead
default count densities can be obtained analytically.

Predictive density for the default rate given default counts up to time
$t-1$ is given by
%
\begin{equation}
\bigl(\theta_{t}|N^{(t-1)}\bigr) \sim \operatorname{Gamma}(\gamma
a_{t-1},\gamma b_{t-1}). \label{thetagnt-1}
\end{equation}
It follows from the above that
$E(\theta_{t}|N^{(t-1)})=E(\theta_{t-1}|N^{(t-1)})$ and\break $V(\theta
_{t}|N^{(t-1)})=\frac{V(\theta_{t-1}|N^{(t-1)})}{\gamma}$.
In other words, the model implies that as we move forward in time, the
expected default rate stays the same but our uncertainty
about the rate increases.

The posterior density of the default rate given default counts up to
time $t$ is given by
%
\begin{equation}
\bigl(\theta_{t}|N^{(t)}\bigr) \sim \operatorname{Gamma}(a_{t},b_{t}),
\label{thetagnt}
\end{equation}
where $a_{t}= \gamma a_{t-1}+N_{t}$ and $b_{t}=\gamma b_{t-1} +1$. The
posterior density (\ref{thetagnt}) is also known as the filtering
distribution of the default rate.\vadjust{\goodbreak}

Finally, one-month-ahead forecasting density of $N_{t}$ given the
default counts up to month $t-1$ can be obtained to be a negative
binomial density as
%
\begin{equation}
\bigl(N_{t}|N^{(t-1)}\bigr) \sim \operatorname{Negbin}(r_{t},
p_{t}), \label{negbin}
\end{equation}
where $r_{t}=\gamma a_{t-1}$ and $p_{t}=\frac{\gamma b_{t-1}}{\gamma
b_{t-1}+1}$. As summarized above, conditional on the discount factor
$\gamma$, the updating of the default rate in light of new default
information and one-month-ahead forecasting densities for default
counts are all available analytically. Another attractive feature of
the proposed model is that in addition to obtaining point estimates of
the default counts and the default rates at each point in time, one can
also obtain well-known probability distributions with easy to obtain
statistical properties such as the mode, median, standard deviation and
credibility intervals.

As noted by the associate editor, ``as homeowners default or pay off
their mortgages, the size of the effective pool of homeowners that
could default changes.'' Thus, knowing the original size of a given
mortgage cohort and the number of people who prepay their mortgages
might have been useful in our analysis. If such information is
available, then it is possible to introduce a Poisson structure that
can capture the behavior of the mortgage size and the prepayment counts
over time by redefining $N_{t}$ as $N_{t} \sim\operatorname{Pois}(h_{t} \lambda
_{t})$ and $P_{t} \sim\operatorname{Pois}(h_{t} \phi_{t})$, where $h_{t}$ is
the mortgage size at time $t$ and $P_{t}$ is the number of prepaid
mortgages at time $t$. Thus, we can keep track of the evolution of
$h_{t}$ as $h_{t} = h_{t-1} - N_{t-1} - P_{t-1}$, where $\lambda_{t}$
and $\phi_{t}$ have Markovian evolutions similar to the one introduced
in (\ref{state1}). Unfortunately, neither the size of the cohorts nor
the prepayment counts were made available to us. In modeling the
default risk at the aggregate level, the size of each cohort is very
large as opposed to monthly default counts as pointed out by \citet{K10}
with the default probability being very small. In our analysis of
mortgage defaults since~$h_{t}$ is not known to us, the default rate
$\theta_{t}$ would be approximately equal to $h_{t}\lambda_{t}$.

\section{Dynamic models with covariates}\label{sec3}
\subsection{Dynamic model with static covariate coefficients}\label{sec3.1}
We next extend the model of Section \ref{sec2} by considering the effects of
covariates on the dynamic default rate. Let $N_{t}$ be the number of
defaults of a given mortgage pool during the month $t$ and $\lambda
_{t}$ be its default rate for $t=1,\ldots,T$. We assume that the
default rate is given by
%
\begin{equation}
\lambda_{t} = \theta_{t}e^{\bolds{\beta}'\mathbf{z}_{t}}, \label{rate}
\end{equation}
where $\mathbf{z}_{t}$ is the vector of the covariates and
$\bolds{\beta}$ is the parameter vector. The covariate vector
$\mathbf{z}_{t}$ may consist of economic variables as well as trend
and seasonal components. Parameter $\theta_{t}$ acts like the baseline
default rate which evolves over time. We also note that (\ref{rate}) is
similar to a proportional hazards model. Given $\lambda_{t}$, we assume
that the number of defaults during the month $t$ is described by a
modulated nonhomogeneous Poisson process,
%
\begin{equation}
(N_{t}|\theta_{t},\bolds{\beta},\mathbf{z}_{t})
\sim \operatorname{Pois}\bigl(\theta_{t}e^{\bolds{\beta}'\mathbf{z}_{t}}\bigr). \label{measurement2}
\end{equation}
The modulated Poisson model (\ref{measurement2}) acts as an observation
equation defined over discrete time. For the state evolution equation
of the baseline failure rate, $\theta_{t}$, we assume the same
structure as before given by (\ref{state1}). In addition, we assume
that initially $(\theta_{0}|\bolds{\beta},\mathbf
{z}_{t},N^{(0)}) \sim \operatorname{Gamma}(a_{0},b_{0})$ and is independent of
$\bolds{\beta}$. Thus, it can be shown that the conditional
distribution of $(\theta_{t-1}|\bolds{\beta},\mathbf
{z}_{t},N^{(t-1)})$ follows a gamma density as
%
\begin{equation}
\bigl(\theta_{t-1}|\bolds{\beta},\mathbf{z}_{t},N^{(t-1)}
\bigr) \sim \operatorname{Gamma}(a_{t-1},b_{t-1}). \label{prior3}
\end{equation}

Therefore, the conditional posterior density of $\theta_{t}$ given
$\bolds{\beta},\mathbf{z}_{t},N^{(t-1)}$ can be obtained via
%
\begin{equation}\qquad
p\bigl(\theta_{t}|\bolds{\beta},\mathbf{z}_{t},N^{(t-1)}
\bigr) = \int^{\infty}_{\gamma\theta_{t}} p\bigl(\theta_{t}|
\theta_{t-1},N^{(t-1)}\bigr)p\bigl(\theta_{t-1}|\bolds{
\beta },\mathbf{z}_{t},N^{(t-1)}\bigr)\,d\theta_{t-1},
\end{equation}
which reduces to a gamma density as
%
\begin{equation}
\bigl(\theta_{t}|\bolds{\beta},\mathbf{z}_{t},N^{(t-1)}
\bigr) \sim \operatorname{Gamma}(\gamma a_{t-1},\gamma b_{t-1}).
\label{thetagnt-1_3}
\end{equation}

Furthermore, the conditional posterior of $\theta_{t}$ given
$\bolds{\beta},\mathbf{z}_{t},N^{(t)}$ can be obtained using
(\ref{measurement2}) and (\ref{thetagnt-1_3}) and the Bayes' rule
%
\begin{equation}
p\bigl(\theta_{t}|\bolds{\beta},\mathbf{z}_{t},N^{(t)}
\bigr) \propto p(N_{t}|\bolds{\beta},\mathbf{z}_{t},
\theta_{t})p\bigl(\theta _{t}|\bolds{\beta},
\mathbf{z}_{t},N^{(t-1)}\bigr).
\end{equation}
The above implies that
\[
p\bigl(\theta_{t}|\bolds{\beta},\mathbf{z}_{t},N^{(t)}
\bigr) \propto \bigl(\theta_{t}e^{\bolds{\beta}'\mathbf{z}_{t}}\bigr)^{\gamma
a_{t-1}+N_{t}-1}e^{-(\gamma
b_{t-1}+1)(\theta_{t}e^{\bolds{\beta}'\mathbf{z}_{t}})},
\]
that is, the conditional distribution of the default rate at time $t$
is a gamma density given by
%
\begin{equation}
\bigl(\theta_{t}|\bolds{\beta},\mathbf{z}_{t},N^{(t)}
\bigr) \sim \operatorname{Gamma}(a_{t},b_{t}), \label{thetagnt4}
\end{equation}
where $a_{t}= \gamma a_{t-1}+ N_{t}$ and $b_{t}=\gamma
b_{t-1} +e^{\bolds{\beta}'\mathbf{z}_{t}}$.

The one-step-ahead conditional predictive distribution of default
counts at time~$t$ given $\bolds{\beta},\mathbf{z}_{t}$ and
$N^{(t-1)}$ can be obtained via
%
\begin{equation}
p\bigl(N_{t}|\bolds{\beta},\mathbf{z}_{t},N^{(t-1)}
\bigr)= \int_{0}^{\infty} p(N_{t}|\bolds{
\beta},\mathbf{z}_{t},\theta _{t})p\bigl(
\theta_{t}|\bolds{\beta},\mathbf {z}_{t},N^{(t-1)}
\bigr)\,d\theta_{t},
\end{equation}
where $(N_{t}|\bolds{\beta},\mathbf{z}_{t},\theta_{t}) \sim
\operatorname{Pois}(\theta_{t}e^{\bolds{\beta}'\mathbf{z}_{t}})$ and $(\theta
_{t}|\bolds{\beta},\mathbf{z}_{t},N^{(t-1)}) \sim \operatorname{Gamma}(\gamma
a_{t-1},\break  \gamma b_{t-1})$. Therefore,
%
\begin{eqnarray}
&& p\bigl(N_{t}|\bolds{\beta},\mathbf{z}_{t},N^{(t-1)}
\bigr)
\nonumber
\\[-8pt]
\\[-8pt]
\nonumber
&&\qquad = \pmatrix{ {\gamma a_{t-1}+N_{t}-1}
\vspace*{2pt}\cr
{N_{t}}}\biggl\{\frac{\gamma b_{t-1}}{\gamma
b_{t-1}+e^{\bolds{\beta}'\mathbf{z}_{t}}}\biggr\}^{\gamma
a_{t-1}}\biggl\{
\frac{e^{\bolds{\beta}'\mathbf
{z_{t}}}}{\gamma b_{t-1}+e^{\bolds{\beta}'\mathbf
{z}_{t}}}\biggr\}^{N_{t}},
\end{eqnarray}
which is a negative binomial model denoted as
%
\begin{equation}
\bigl(N_{t}|N^{(t-1)}, \bolds{\beta},\mathbf{z}_{t}
\bigr) \sim \operatorname{Negbin}(r_{t}, p_{t}), \label{negbin2}
\end{equation}
where $r_{t}=\gamma a_{t-1}$ and $p_{t}=\frac{\gamma b_{t-1}}{\gamma
b_{t-1}+e^{\bolds{\beta}'\mathbf{z}_{t}}}$. The predictive
density (\ref{negbin2}) implies that given the covariates and the
default counts up to month $t-1$, forecasts for the month $t$ are a
function of the observed default count in month $t-1$ adjusted by the
corresponding covariates. The mean of $(N_{t}|N^{(t-1)},\bolds
{\beta},\mathbf{z}_{t})$ can be computed via
%
\begin{equation}
E\bigl(N_{t}|N^{(t-1)},\bolds{\beta},\mathbf{z}_{t}
\bigr) = \frac
{a_{t-1}}{b_{t-1}} e^{\bolds{\beta}'\mathbf{z}_{t}}. \label{meannbin}
\end{equation}

Since the results previously presented are conditional on the parameter
vector~$\bolds{\beta}$ and the discount factor $\gamma$, we next
discuss how to obtain the posterior distributions of $\bolds{\beta
}$ and $\gamma$. Since these distributions cannot be obtained
analytically, we will use Markov chain Monte Carlo (MCMC) methods to
generate samples from these posterior distributions.

\subsection{Dynamic model with dynamic covariate coefficients}\label{sec3.2}
A natural extension of the dynamic model with covariates is to let the
regression coefficients vary over time. As pointed out by one of the
reviewers, in volatile economic environments, macroeconomic variables
might exhibit sudden ups and downs and being able to capture the
effects of such changes will be of concern to institutions that are
managing the mortgage loans. One way to take into account such changes
is to allow a dynamic structure on the covariate coefficients.
Following the same notation introduced previously, we assume that the
number of defaults during the month $t$ is described by
%
\begin{equation}
(N_{t}|\theta_{t},\bolds{\beta}_{t},
\mathbf{z}_{t}) \sim \operatorname{Pois}\bigl(\theta_{t}e^{\bolds{\beta_{t}}'\mathbf{z}_{t}}
\bigr). \label{measurement3}
\end{equation}
In (\ref{measurement3}), $\bolds{\beta}_{t}$s are time varying
coefficients and we assume the same structure on $\theta_{t}$ as in
(\ref{state1}). The conditional updating of $\theta_{t}$s will be the
same as introduced in Section~\ref{sec3} and their details will be omitted from
the discussion to preserve space. Time evolution of the covariate
coefficients is described by
%
\begin{equation}
\beta_{it} \sim N(\beta_{it-1}, \tau_{i})\qquad \forall
i,
\end{equation}
where $i$ represents the covariate index, $\tau_{i}$ is the precision
parameter for each $i$ and its prior is assumed to be
%
\begin{equation}
\tau_{i} \sim \operatorname{Gamma}(a_{\tau}, b_{\tau}) \qquad\forall i.
\end{equation}
We use this extension in our numerical example in Section \ref{sec5} to learn if
dynamic nature of the covariate effects improves model fit and
forecasting performance.

\section{Bayesian analysis}\label{sec4}
Most of the parameter updating and forecasting for the dynamic model
presented in Section \ref{sec2} is available in closed form given that the
discounting term $\gamma$ is known.\vadjust{\goodbreak} Alternatively, one can assume an
unknown $\gamma$ and use Bayesian analysis to carry out inference.
Following the development of Section~\ref{sec3}, the distributions obtained for
the dynamic model with covariates are all conditional on $\bolds
{\beta}$ and $\gamma$. Our objective is to obtain the posterior joint
distribution of the model parameters given that we have observed all
default counts up to time $t$, that is, $p(\theta_{1},\ldots,\theta
_{t}|N^{(t)})$ for the dynamic model and $p(\theta_{1},\ldots,\theta
_{t}, \bolds{\beta}|N^{(t)})$ for the dynamic model with
covariates, both of which can be used to infer mortgage default risk
behavior of a given cohort. In addition, being able to obtain
one-month-ahead predictive distributions of the default counts,
$p(N_{t}|N^{(t-1)})$, will be of interest to institutions that are
managing the loans.

\subsection{Posterior inference}\label{sec4.1}
Since our goal is to obtain $p(\theta_{1},\ldots,\theta_{t}, \bolds
{\beta}|N^{(t)})$ which is not available in closed form, we can use a
Gibbs sampler to generate samples from it. In order to do so, we need
to be able to generate samples from the full conditional distributions
of $p(\theta_{1},\ldots,\theta_{t}|\bolds{\beta},N^{(t)})$ and
$p(\bolds{\beta}| \theta_{1},\ldots,\theta_{t}, N^{(t)})$, none of
which are available as known densities. Next, we discuss how to
generate samples from these densities.

The conditional posterior distribution of $\bolds{\beta}$ given
the default rates can be obtained by
%
\begin{equation}
p\bigl(\bolds{\beta}|\theta_{1},\ldots,\theta_{t},z_{t},N^{(t)}
\bigr) \propto\prod_{i=1}^{t}
\frac{\operatorname{exp} \{\theta_{i}e^{\bolds{\beta
}'\mathbf{z}_{i}}\}(\theta_{i}e^{\bolds{\beta
}'\mathbf{z}_{i}})^{N_{i}}}{N_{i}!}p(\bolds{\beta}), \label{betafullcon}
\end{equation}
where $p(\bolds{\beta})$ is the prior for $\bolds{\beta}$.
Regardless of the prior selection for $\bolds{\beta}$, (\ref
{betafullcon}) will not be a known density. Therefore, we can use a
random walk Metropolis--Hastings algorithm to be able to generate
samples from $p(\bolds{\beta}|\theta_{1},\ldots,\theta
_{t},z_{t},N^{(t)})$. Following \citet{CG95}, the steps in the
Metropolis--Hastings algorithm can be summarized as follows:
\begin{longlist}[1.]
\item[1.] Assume the starting points $\bolds{\beta}^{(0)}$ at $j=0$.

Repeat for $j>0$.
\item[2.] Generate $\bolds{\beta}^{*}$ from $q(\bolds{\beta
}^{*}|\bolds{\beta}^{(j)})$ and $u$ from $U(0,1)$.
\item[3.] If $u \le a(\bolds{\beta}^{(j)},\bolds{\beta}^{*})$
then set $\bolds{\beta}^{(j)}=\bolds{\beta}^{*}$; else set
$\bolds{\beta}^{(j)}=\bolds{\beta}^{(j)}$ and $j=j+1$,
where
%
\begin{equation}
a\bigl(\bolds{\beta}^{(j)},\bolds{\beta}^{*}\bigr) = \operatorname{min}
\biggl\{ 1,\frac{\pi(\bolds{\beta}^{*})q(\bolds{\beta
}^{(j)}|\bolds{\beta}^{*})}{\pi(\bolds{\beta
}^{(j)})q(\bolds{\beta}^{*}|\bolds{\beta}^{(j)})} \biggr\}. \label{accept}
\end{equation}
\end{longlist}

In (\ref{accept}), $\pi(\cdot)$ is given by (\ref{betafullcon}), that is,
the density we need to generate samples from, and $q(\cdot|\cdot)$ is the
multivariate normal proposal density whose variance-covariance matrix
is determined via $(-H)^{-1}$ with $H$ representing the approximate
Hessian of $\pi(\cdot)$ evaluated at its mode; see \citet{gelman}. If we
repeat the above a large number of times, then we obtain samples from
$p(\bolds{\beta}|\theta_{1},\ldots,\theta_{t},z_{t},N^{(t)})$.
Next, we discuss how one can generate samples from the other full
conditional distribution, $p(\theta_{1},\ldots, \theta_{t}|\bolds
{\beta},\mathbf{z}_{t},N^{(t)})$.

Due to the Markovian nature of the default rates, using the chain rule,
we can rewrite the full conditional density, $p(\theta_{1},\ldots,\theta
_{t}|\bolds{\beta},\mathbf{z}_{t}, N^{(t)})$, as
%
\begin{equation}
p\bigl(\theta_{t}|\bolds{\beta},\mathbf{z}_{t},
N^{(t)}\bigr)p\bigl(\theta _{t-1}|\theta_{t}, \bolds{
\beta},\mathbf{z}_{t}, N^{(t-1)}\bigr) \cdots p\bigl(
\theta_{1}|\theta_{2}, \bolds{\beta },\mathbf{z}_{t},
N^{(1)}\bigr). \label{jointtheta}
\end{equation}
In (\ref{jointtheta}), $p(\theta_{t}|\bolds{\beta},\mathbf
{z}_{t}, N^{(t)})$ is available from
(\ref{thetagnt4}) and $p(\theta_{n-1}|\theta_{n}, \bolds{\beta
},\mathbf{z}_{t},
N^{(n-1)})$ for any $n$ can be obtained as follows:
%
\begin{eqnarray} \label{forwback}
&&p\bigl(\theta_{n-1}|\theta_{n}, \bolds{\beta},
\mathbf{z}_{t}, N^{(n-1)}\bigr)
\nonumber
\\[-8pt]
\\[-8pt]
\nonumber
&&\qquad\propto p\bigl(
\theta_{n}|\theta_{n-1}, \bolds{\beta },\mathbf{z}_{t},
N^{(n-1)}\bigr)p\bigl(\theta_{n-1}| \bolds{\beta },
\mathbf{z}_{t}, N^{(n-1)}\bigr).
\end{eqnarray}
It can be shown that $(\theta_{n-1}|\theta_{n}, \bolds{\beta
},\mathbf{z}_{t}, N^{(n-1)}) \sim \operatorname{Gamma}[(1-\gamma) a_{n-1},
b_{n-1}]$, where $\gamma\theta_{n} < \theta_{n-1} < \infty$, that is, a
truncated gamma density.

Therefore, given (\ref{jointtheta}) and the posterior samples generated
from the full conditional distribution of $\bolds{\beta}$, we can
sample from $p(\theta_{1},\ldots,\theta_{t}|\bolds{\beta
},\mathbf{z}_{t}, N^{(t)})$ by sequentially simulating the
individual default rates as follows:
\begin{longlist}[1.]
\item[1.] Assume the starting points $\theta_{1}^{(0)},\ldots, \theta
_{t}^{(0)}$ at $j=0$.

Repeat for $j>0$.
\item[2.] Using the generated $\bolds{\beta}^{(j)}$, sample $\theta
^{(j)}_{t}$ from $(\theta_{t}|\bolds{\beta}^{(j)},\mathbf
{z}_{t}, N^{(t)})$.
\item[3.] Using the generated $\bolds{\beta}^{(j)}$, for each $n =
t-1, \ldots,1$ generate $\theta^{(j)}_{n}$ from $(\theta_{n}|\theta
^{(j)}_{n+1}, \bolds{\beta},\mathbf{z}_{t}, N^{(n)})$ where
$\theta^{(j)}_{n+1}$ is the value generated in the previous step.
\end{longlist}
If we repeat the above a large number of times, then we obtain samples
from the joint full conditional distribution of default rates. The
generation of $\theta^{(j)}_{n}$s in step 3 above is known as the
\textit{forward filtering backward sampling algorithm}; see \citet
{FS94}. Consequently, we can obtain samples from the joint density of
the model parameters by iteratively sampling from $p(\bolds{\beta
}|\theta_{1},\ldots,\theta_{t},z_{t},N^{(t)})$ and $p(\theta_{1},\ldots,\theta_{t}|\bolds{\beta},\mathbf{z}_{t}, N^{(t)})$, namely, a
full Gibbs sampler algorithm; see \citet{SG92}.

The FFBS algorithm as discussed above can also be used to generate
samples from $p(\theta_{1},\ldots,\theta_{t}| N^{(t)})$ for the dynamic
model without the use of the additional Gibbs sampler step for
$\bolds{\beta}$. In addition, the above algorithm allows us to
obtain a density estimate for $p(\theta_{t-k}| N^{(t)})$ for all $k
\geq1$ for both dynamic models which can be used for retrospective
comparison of default rates among different mortgage pools. To the best
of our knowledge, this type of approach has not been considered in the
mortgage default risk literature.

\subsection{Unknown discount parameter $\gamma$}\label{sec4.2}
Previously the discount factor $\gamma$ has been assumed to be known.
If $\gamma$ were to be treated as an unknown quantity, then it is
possible to obtain its Bayesian updating. Following the development of
the dynamic model introduced in Section \ref{sec2}, the posterior distribution
of $\gamma$ can be obtained by
%
\begin{equation}
p\bigl(\gamma|N^{(t)}\bigr) \propto\prod_{k=1}^{t}
p\bigl(N_{k}|N^{(k-1)},\gamma\bigr)p(\gamma),
\label{gammadispost2}
\end{equation}
where $p(N_{k}|N^{(k-1)},\gamma)$ is the likelihood term which is given
by (\ref{negbin}) and $p(\gamma)$ is the prior for $\gamma$. Since (\ref
{gammadispost2})
will not be a known density for any prior for $\gamma$, we need to
sample from the posterior distribution of $\gamma$ using MCMC. As an
alternative, a~discrete prior over $(0,1)$ can be considered which can
numerically be summed out from (\ref{gammadispost2}).

For the dynamic model with covariates detailed in Section \ref{sec3}, one can
generate samples from the posterior joint distribution of $\gamma$ and
$\bolds{\beta}$ from the following:
%
\begin{equation}
p\bigl(\gamma, \bolds{\beta}|N^{(t)},\mathbf{z}_{t}\bigr)
\propto p(N_{1},\ldots,N_{t}|\mathbf{z}_{t},
\gamma, \bolds{\beta })p(\gamma, \bolds{\beta}), \label{gammabetajoint}
\end{equation}
where $p(\gamma, \bolds{\beta})=p(\gamma)p(\bolds{\beta})$
when $\gamma$ and $\bolds{\beta}$ are assumed to be independent a
priori and the likelihood term, $p(N_{1},\ldots,N_{t}|\mathbf
{z}_{t},\gamma, \bolds{\beta})$, can be obtained as
%
\begin{equation}
\qquad p(N_{1},\ldots,N_{t}|\mathbf{z}_{t},\gamma,
\bolds{\beta}) = L\bigl(\gamma, \bolds{\beta};\mathbf{z}_{t},N^{(t)}
\bigr)= \prod_{k=1}^{t} p
\bigl(N_{k}|N^{(k-1)},\mathbf{z}_{t},\bolds{\beta },
\gamma\bigr), \label{jointlike}
\end{equation}
where $p(N_{k}|N^{(k-1)},\mathbf{z}_{t},\bolds{\beta},\gamma)$
is given by (\ref{negbin2}). The fact that (\ref{jointlike}) is free of
$\theta_{t}$s facilitates the posterior generation. Since (\ref
{gammabetajoint}) will not be available in closed form for any prior of
$\gamma$ and $\bolds{\beta}$, one can use a Metropolis--Hastings
algorithm to generate samples from the joint posterior density as
presented in Section~\ref{sec3.1}. This approach can also be used to estimate
$p(\gamma, \bolds{\beta}_{t}|\tau_{1},\ldots,\tau
_{2},N^{(t)},\mathbf{z}_{t})$
for the model with dynamic covariates of Section \ref{sec4.1}. Thus, a Gibbs
sampler can be used to obtain samples from the full joint distribution
of all model parameters by iteratively generating samples between
$p(\gamma, \bolds{\beta}_{t}|\tau_{1},\ldots,\tau
_{2},N^{(t)},\mathbf{z}_{t})$ and $p(\tau_{i}|\bolds{\beta
}_{t},\gamma,N^{(t)},\mathbf{z}_{t})$'s.

In addition, $p(\theta_{1},\ldots,\theta_{t}|N^{(t)},\mathbf
{z}_{t},\bolds{\beta},\gamma)$, the conditional joint distribution
of the default rates, can be obtained using the FFBS algorithm as
presented in Section~\ref{sec4.1}. Thus, the joint smoothing distribution of the
default rates can be computed by
%
\begin{equation}
\qquad p\bigl(\theta_{1},\ldots,\theta_{t}|N^{(t)}\bigr)
= \int\!\!\int p\bigl(\theta_{1},\ldots,\theta_{t}|N^{(t)},
\mathbf{z}_{t},\bolds{\beta},\gamma \bigr)p\bigl(\gamma,\bolds{\beta} |
N^{(t)}\bigr) \,d\gamma \,d\bolds{\beta}, \label{jointsmt}
\end{equation}
where only samples from $p(\gamma,\bolds{\beta}|N^{(t)})$ will be
available. Therefore, the above can be approximated as a Monte Carlo
average via
%
\begin{equation}
p\bigl(\theta_{1},\ldots,\theta_{t}|N^{(t)}\bigr)
\approx \frac{1}{S} \sum_{j=1}^{S} p
\bigl(\theta_{1},\ldots,\theta_{t}|N^{(t)},\mathbf
{z}_{t},\bolds{\beta}^{(j)},\gamma^{(j)}\bigr),
\label{jointsmt2}
\end{equation}
where $S$ is the number of samples, and $(\bolds{\beta}^{(j)},
\gamma^{(j)})$ are the generated sample pairs.

\subsection{One-month-ahead forecasting}\label{sec4.3}
In order to obtain one-month-ahead forecast distributions from the
dynamic model with covariates, the\vadjust{\goodbreak} following can be used:
%
\begin{equation}
p\bigl(N_{t}|N^{(t-1)},\mathbf{z}_{t}\bigr) = \int\!\!
\int p\bigl(N_{t}|N^{(t-1)},\mathbf{z}_{t},\bolds{
\beta},\gamma\bigr)p\bigl(\gamma, \bolds{\beta}|N^{(t)}\bigr) \,d \bolds{
\beta} \,d \gamma. \label{forecastdmwc}
\end{equation}
Since only samples from $p(\gamma, \bolds{\beta}|N^{(t)})$ will be
available, the above can be approximated by
%
\begin{equation}
p\bigl(N_{t}|N^{(t-1)},\mathbf{z}_{t}\bigr) \approx
\frac{1}{S} \sum_{j=1}^{S} p
\bigl(N_{t}|N^{(t-1)},\mathbf{z}_{t},\bolds{\beta
}^{(j)},\gamma^{(j)}\bigr). \label{forecastdmwc2}
\end{equation}
Similarly, (\ref{forecastdmwc2}) can be computed for the dynamic model
of Section \ref{sec2} without any covariates and the dynamic model with dynamic
covariates of Section~\ref{sec4.1}.

\subsection{Model comparison}\label{sec4.4}
In order to compare the fit of the proposed models to data, we consider
two sets of measures that are used with sampling based methods, the
Bayes factor with the harmonic mean estimator and the pseudo Bayes
factor with the conditional predictive ordinate. In what follows, we
briefly summarize both methods whose implementations are discussed in
our numerical example.

\subsubsection{Bayes factor-harmonic mean estimator}\label{sec4.4.1}
The first fit measure is the Bayes factor approximation of models with
MCMC steps; we refer to this measure as the Bayes factor-harmonic mean
estimator which has been discussed by \citet{GDC92} and \citet{KR95}. The
harmonic mean estimator of the predictive likelihood for a given model
can be obtained as
%
\begin{equation}
p\bigl(N^{(t)}\bigr) = \Biggl\{\frac{1}{S} \sum
_{j=1}^{S} p\bigl(N^{(t)}|\bolds {
\Theta}^{(j)}\bigr)^{-1} \Biggr\}^{-1},
\label{bfmcmc}
\end{equation}
where $S$ is the number of iterations and $\bolds{\Theta}^{(j)}$
is the $j$th generated posterior sample. For the proposed models, (\ref
{bfmcmc}) can be computed via
%
\begin{equation}
p\bigl(N^{(t)}\bigr) = \Biggl\{\frac{1}{S} \sum
_{j=1}^{S} \Biggl\{\prod
_{k=1}^{t} p\bigl(N_{k}|N^{(k-1)},
\bolds{\Theta}^{(j)}\bigr) \Biggr\}^{-1} \Biggr
\}^{-1}, \label{bfmcmc2}
\end{equation}
where $p(N_{k}|N^{(k-1)}, \bolds{\Theta}^{(j)})=p(N_{k}|N^{(k-1)},
\gamma^{(j)})$ can be obtained via (\ref{negbin}) and
$p(N_{k}|N^{(k-1)}, \bolds{\Theta}^{(j)})=p(N_{k}|N^{(k-1)},
\mathbf{z}_{t},\bolds{\beta}^{(j)},\gamma^{(j)})$ via (\ref
{negbin2}). In comparing two models, a higher $p(N^{(t)})$ value
indicates a better fit. As pointed out by \citet{KR95}, although the use
of (\ref{bfmcmc}) has been criticized due to potential large effects of
a sample value on the likelihood, it has been shown to give accurate
results in most cases and is preferred for its computational simplicity.

\subsubsection{Pseudo Bayes factor-conditional predictive ordinate}\label{sec4.4.2}
An alternative method to compare models with sampling based estimation
is the calculation of the pseudo Bayes factor using the conditional
predictive ordinate. Following \citet{G96}, the comparison criteria
makes use of a cross-validation estimate of the marginal likelihood.
The main advantage of this approach is once again its computational simplicity.

The cross validation predictive density for the $i$th observation is
defined as $f(N_{i}|\mathbf{N}^{(-i)})$, where $\mathbf{N}^{(-i)}$
represents the data, $N^{(i)}$, except for $N_{i}$ and can be estimated via
%
\begin{equation}
\hat{f}\bigl(N_{i}|\mathbf{N}^{(-i)}\bigr)=
\frac{1}{{1}/{S} \sum_{j=1}^{S}
{1}/{(f(N_{i}|\mathbf{N}^{(-i)},\bolds{\Theta}^{(j)}))} }, \label{monteestimate}
\end{equation}
where $S$ is the number of samples generated and $\bolds{\Theta
}^{(j)}$ is the $j$th generated parameter sample vector. Since given
$\bolds{\Theta}$, $N_{i}$s are independent, $f(N_{i}|\mathbf
{N}^{(-i)},\break\bolds{\Theta}^{(j)})=f(N_{i}|\bolds{\Theta
}^{(j)})$ can be used in (\ref{monteestimate}). Once the
cross-validation predictive densities are estimated using (\ref
{monteestimate}), one can compare the proposed models in terms of fit
in the log-scale. In comparing models, a higher conditional predictive
ordinate indicates a better fit.

\subsubsection{A~Bayesian Poisson regression and an EWMA as benchmark models}\label{sec4.4.3}
A~Bayesian Poisson regression model can be used to test the dynamic
nature of the default rate and also can act as a benchmark model for an
out-of-sample forecasting exercise. In this case, we assume that the
default counts, $N_{t}$'s, follow a nonhomogeneous Poisson process whose
default rate is $\theta_t$ where $\theta_t = \operatorname{exp} \{\bolds
{\beta}'\mathbf{z}_{t} \}$, which can be obtained as a special
case of the model in Section \ref{sec3.1} if $\gamma=1$ in the state evolution
of $\theta_{t} =\frac{\theta_{t-1}}{\gamma}\varepsilon_{t}$. In other
words, the default rate is a deterministic function of the covariates
and is not stochastically evolving over time unlike the dynamic models.
In order to obtain the posterior distribution of the model parameters,
$\bolds{\beta}$, we can use the Metropolis--Hastings algorithm as
discussed in Section \ref{sec4.1}, where the likelihood function is given by
%
\begin{equation}
L\bigl(\bolds{\beta};N^{(t)},\mathbf{z}_{t}\bigr) = \prod
_{i=1}^{t} \frac
{\operatorname{exp} \{ e^{\bolds{\beta}'\mathbf{z}_{i}}\}
(e^{\bolds{\beta}'\mathbf{z}_{i}})^{N_{i}}}{N_{i}!},
\label{poisreglik}
\end{equation}
and each $\beta$ coefficient is a priori, assumed to be normally distributed.

In addition, we also consider using a simple time series model such as
the exponentially weighted moving average (EWMA) to test the
forecasting performance of the proposed models. To determine the
smoothing constant (say, $\nu$) for the EWMA model, we sequentially
minimized the mean absolute percentage deviations and estimated it each
time to predict the next month's default counts using the following:
\[
\hat{N}_{t+1} = \nu N_{t} + (1-\nu) \hat{N}_{t},
\]
where $\hat{N}_{t}$ represents the prediction for month $t$ given
observations up to month $t-1$. Thus, the mean absolute percentage to
be minimized can be written as $\frac{1}{T}\sum_{t=1}^{T} |N_{t}-\hat{N}_{t}|$.

\section{Numerical analysis of monthly mortgage default counts}\label{sec5}
\subsection{Description of default data}\label{sec5.1}
In order to illustrate how the proposed models can be applied to real
mortgage default risk, we used the data provided by the Federal Housing
Administration (FHA) of the U.S. Department of Housing and Urban
Development (HUD). The data consists of defaulted FHA insured single
family mortgage loans originated in different years and in four regions
where HUD has local offices. In our analysis of the default counts, we
use a subset of the data which consists of monthly defaulted FHA
insured single-family 30-year fixed rate (30-yr FRM) mortgage loans
between the dates of January 1994 and December 2005 in the Atlanta
region. We refer to this cohort as the 1994 cohort in the narrative.

Since default behavior is influenced by factors relating to both the
housing equity and the mortgage borrower's ability to pay the loan, we
consider two equity and two ability-to-pay covariates in our analysis.
Housing equity is mainly determined by the housing price level and
interest rate. Therefore, we include the regional conventional mortgage
home price index (CMHPI) and the federal cost of funds index (COFI) as
aggregate equity factors. The CMHPI and COFI are provided by Freddie
Mac and are used as benchmark indices in the U.S. residential mortgage
market. In addition, in order to take into account borrowers' overall
repayment ability, we consider the homeowner mortgage financial
obligations ratio (FOR Mortgage) from The Federal Reserve Board which
reflects periodical mortgage repayment burden of borrowers, and
regional unemployment rate from the U.S. Census, which represents the
impact from trigger events at the aggregate level.

\begin{figure}

\includegraphics{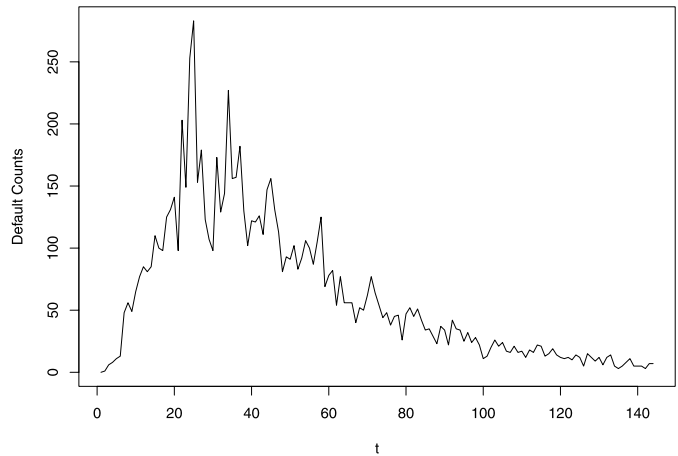}

\caption{Monthly default counts between January 1994 and December 2005
for the 1994 Cohort.} \label{defaultcount}
\end{figure}

As seen in Figure \ref{defaultcount} between January 1994 and December
2005, the default counts for the 1994 cohort seem to exhibit a
nonstationary behavior which can be captured by our state space models.
In what follows, we illustrate the implementation of each model,
discuss implications and present fit measures.

\begin{figure}

\includegraphics{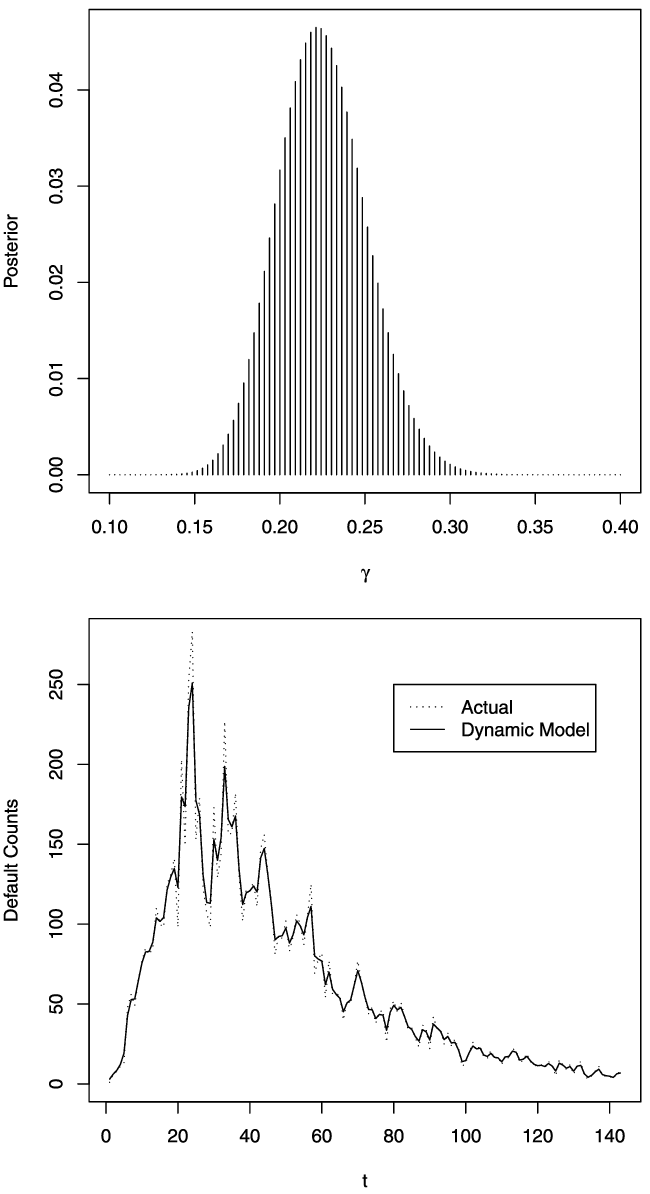}

\caption{Posterior $\gamma$ of the dynamic model (top) and the
retrospective fit of the dynamic model to data (bottom).} \label{gamma1}
\end{figure}

\subsection{\texorpdfstring{Analysis of dynamic model of Section \protect\ref{sec2}}
{Analysis of dynamic model of Section 2}}\label{sec5.2}
As discussed in Section \ref{sec2}, the dynamic model assumes that the default
counts are observations from a nonhomogeneous Poisson process whose
rate is stochastically evolving over time. The attractive feature of
the dynamic model with no covariates is its analytical tractability and
straight forward updating scheme. In our analysis, we assumed that the
discounting factor $\gamma$ given in (\ref{state1}) follows a discrete
uniform distribution defined over $(0,1)$ by the hundredths place and
obtained its posterior density via (\ref{gammadispost2}). As shown in
Figure \ref{gamma1}, the posterior distribution of $\gamma$ is
concentrated around 0.15 and 0.32 with a mean of 0.23. Using the
posterior of $\gamma$ and the forward filtering backward sampling
algorithm presented in Section \ref{sec5.1}, one can obtain the retrospective
fit of the default rate given data. An overlay plot of the mean
posterior default rate and the actual data is shown in Figure \ref
{gamma1}, where evidence in favor of the proposed dynamic model can be
inferred. Given the joint distribution of the default rate over time,
that is, $p(\theta_{1},\ldots,\theta_{t}|N^{(t)})$, the financial
institution managing the loans will have a better understanding of the
default behavior of a given cohort and can use it to manage risk or
explain potential behavior of similar cohorts. In addition, Bayesian
analysis of the mortgage default risk allows direct comparison of the
default rates during different time periods probabilistically. For
instance, one can compute the posterior probability that the default
rate during the second month is greater than that of the first month
for a given cohort. For example, $p(\theta_{2} \ge\theta_{1}|N^{(t)})$
was computed to be 0.3387.

\subsection{\texorpdfstring{Analysis of dynamic models with static and dynamic
covariate coefficients of Section \protect\ref{sec3}}
{Analysis of dynamic models with static and dynamic
covariate coefficients of Section 3}}
\begin{figure}

\includegraphics{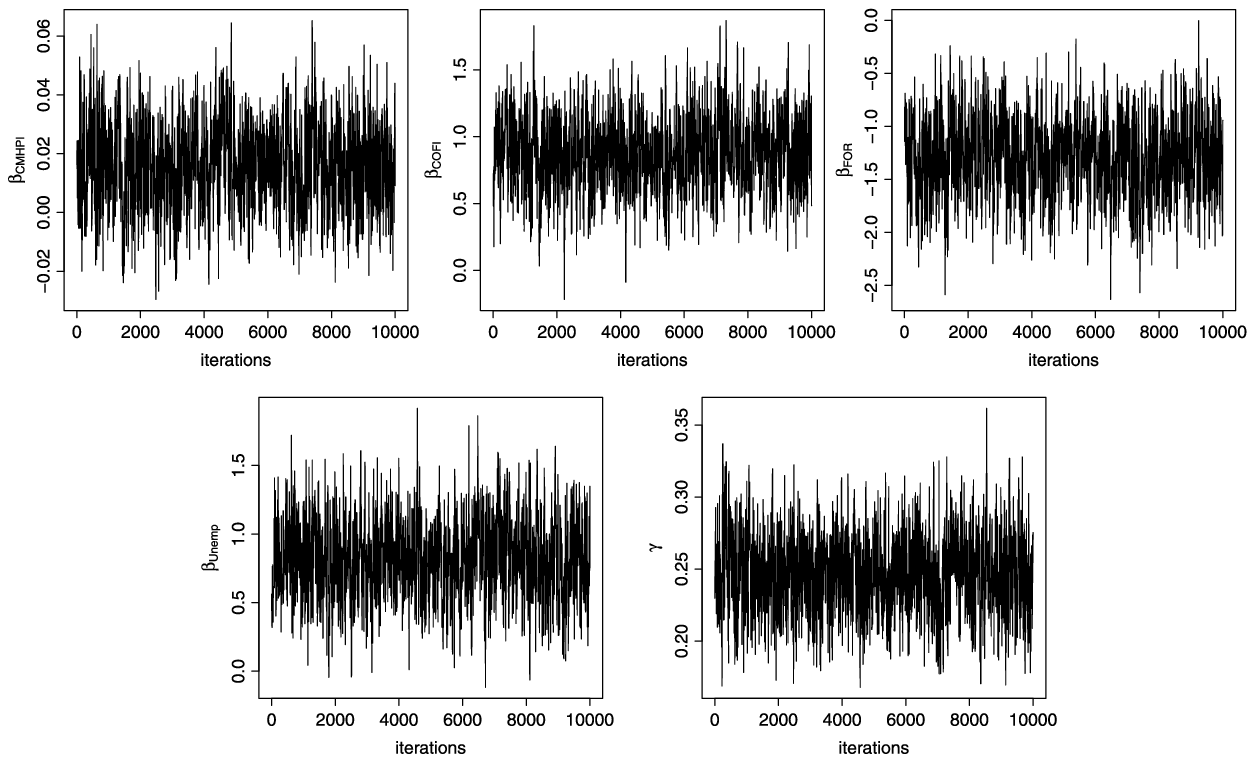}

\caption{Trace plots of $\bolds{\beta}$ and $\gamma$ of the
dynamic model with covariates.} \label{betatrace}
\end{figure}

\begin{figure}[b]\vspace*{-3pt}

\includegraphics{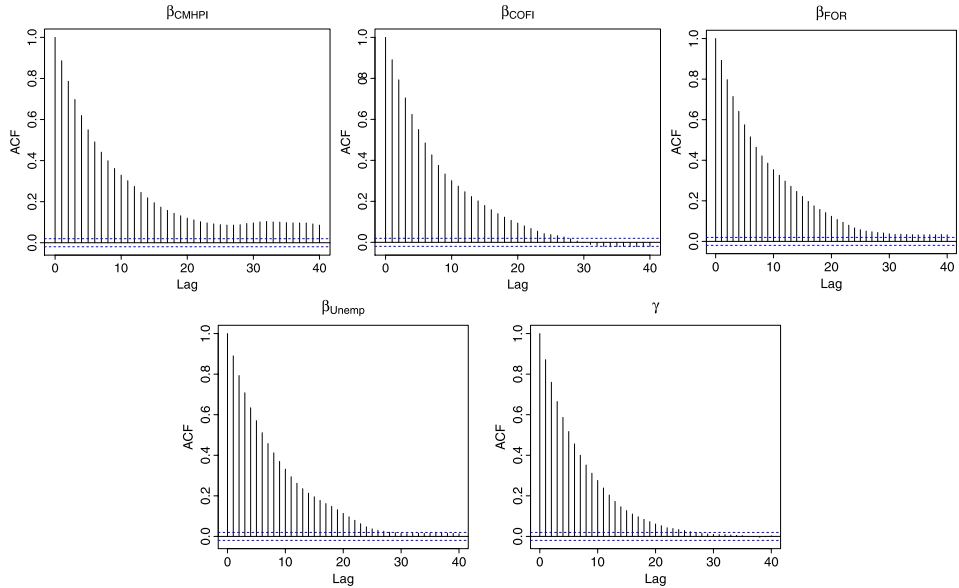}

\caption{Autocorrelation plots of $\bolds{\beta}$ and $\gamma$ of
the dynamic model with covariates.} \label{betaacf}
\end{figure}

In taking into account the effects of macroeconomic variables, we
estimated the dynamic model with covariates as presented in Section \ref{sec3}.
In doing so, we assumed flat but proper priors for the model
parameters. More specifically, the discounting term, $\gamma$, a priori
follows a continuous uniform distribution defined over $(0,1)$ and the
covariate coefficients, $\bolds{\beta}$, follow independent normal
distributions as $\beta_{i} \sim N(0,100)\ \forall i$. In addition,
we also estimated the model using a Beta prior on $\gamma$ as $\gamma
\sim B(3,3)$ to assess prior sensitivity as suggested by one of the
referees and the results were identical. We ran the MCMC algorithm for
10,000 iterations with a burn-in period of 2000 iterations, and did
not encounter any convergence issues. The trace plots for the posterior
samples are shown in Figure \ref{betatrace} and their autocorrelation
plots are shown in Figure \ref{betaacf}, both of which informally show
support in favor of convergence.

\begin{figure}

\includegraphics{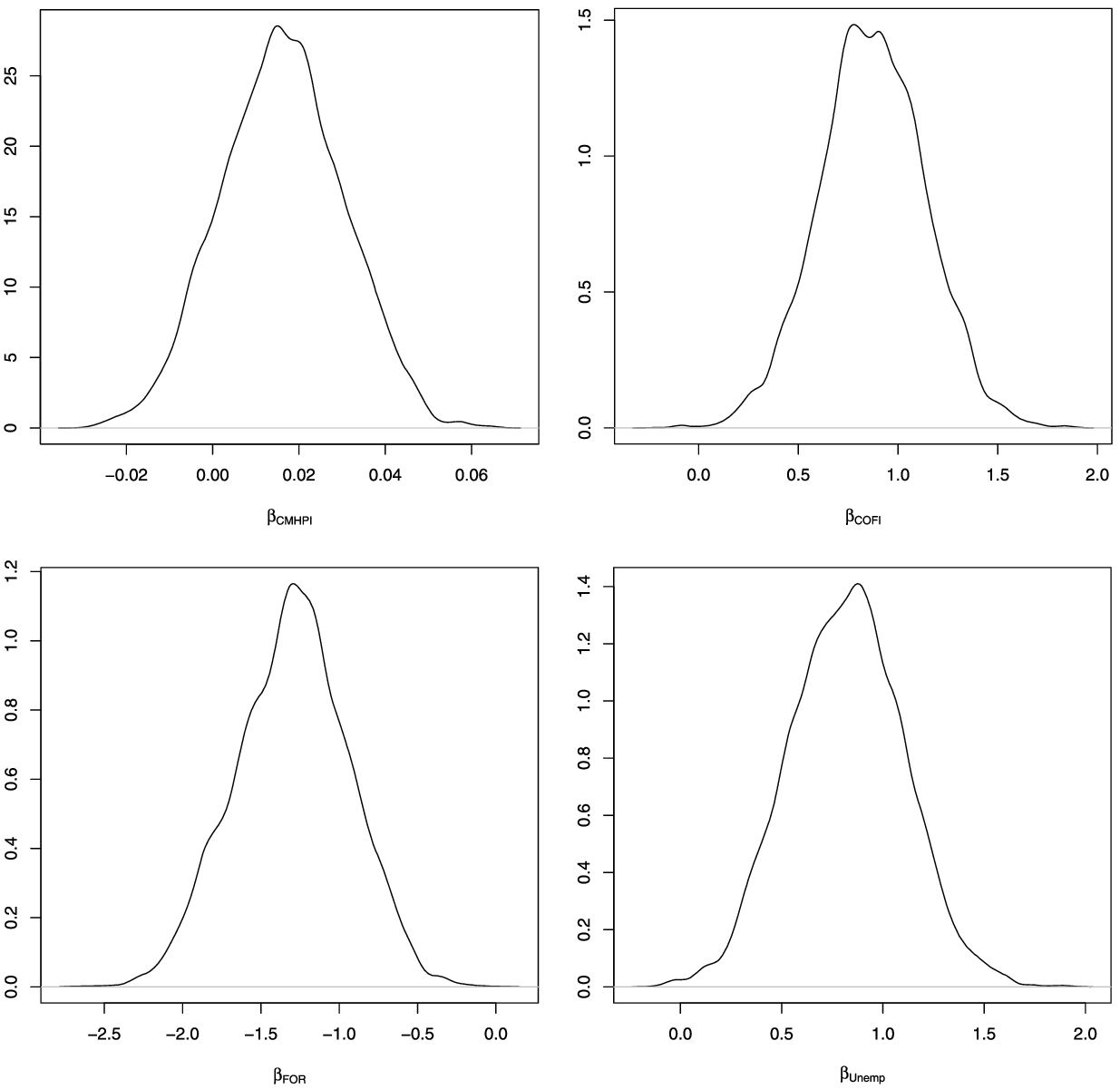}

\caption{Posterior density plots of $\bolds{\beta}$ of the dynamic
model with covariates.} \label{betas}
\end{figure}

\begin{figure}

\includegraphics{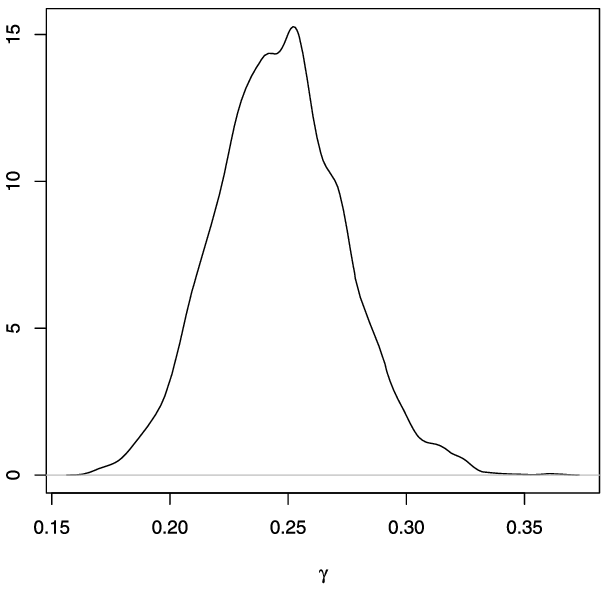}

\caption{Posterior density plot of $\gamma$ of the dynamic model with
covariates.} \label{gamma2}\vspace*{-3pt}
\end{figure}

The posterior density plots of $\bolds{\beta}$ are shown in Figure
\ref{betas} and of $\gamma$ in Figure~\ref{gamma2}, which exhibits
similar behavior to that of the posterior discounting term obtained for
the dynamic model as in Figure \ref{gamma1}.

As can be observed from Table \ref{summary1}, the $\beta$ coefficients
all have fairly significant effects on the default rate. An advantage
of the Bayesian approach is its ability to quantify posterior inference
probabilistically. For instance, one can calculate the probability that
$\beta_{\mathrm{CMHPI}}$ is greater than 0, that is, $p(\beta
_{\mathrm{CMHPI}}>0|N^{(t)})$. Given the cohort at hand, $P(\beta
_{\mathrm{CMHPI}}>0|N^{(t)})$ was obtained to be approximately 0.87, which shows
strong evidence in favor of a positive effect. In summary, the regional
conventional mortgage home price index (CMHPI), federal cost of funds
index (COFI) and the regional unemployment rate (Unemp) have positive
effects on default counts. For instance, as unemployment goes up, the
model suggests that the number of people defaulting tends to increase
for the cohort under study. On the other hand, the homeowner financial
obligations ratio (FOR) seems to decrease the expected number of
defaults as it goes up, namely, as the burden of repayment becomes
relatively easier, then homeowners are less likely to default.

\begin{figure}

\includegraphics{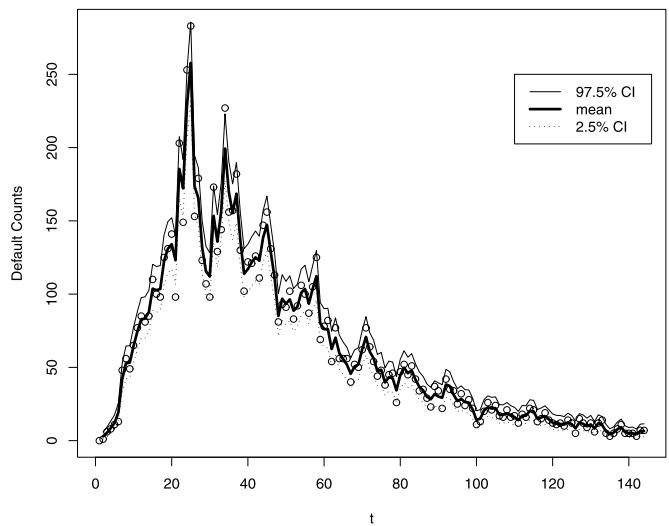}

\caption{Retrospective fit of the dynamic model with static covariate
coefficients to data.} \label{fitpreg}\vspace*{-3pt}
\end{figure}

\begin{figure}[b]\vspace*{-3pt}

\includegraphics{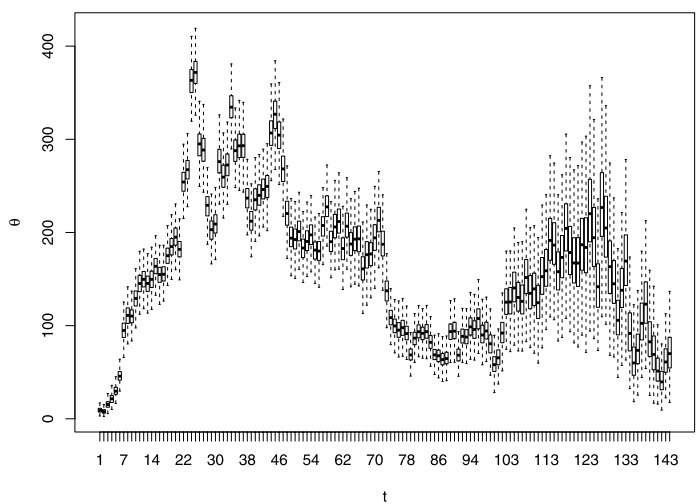}

\caption{Boxplots for smoothed $\theta_{t}$s from $p(\theta_{1},\ldots,\theta_{t}|N^{(t)})$.} \label{smtheta}
\end{figure}

One of the issues that had been under investigation so far was the
dynamic nature of the default rate. As shown in Figure \ref{fitpreg},
the fit of the dynamic model with covariates is reasonably good,
justifying the dynamic behavior of the default rate. Similar
conclusions can be drawn for the dynamic model without the covariates
whose fit is shown in the right panel of Figure \ref{gamma1}. In
showing the dynamic nature of the default rate, we obtained the joint
distribution of the baseline default rates, that is, $p(\theta
_{1},\ldots,\theta_{t}|N^{(t)})$ as in (\ref{jointsmt2}). A boxplot of
$\theta_{t}$s is shown in Figure \ref{smtheta}, which once again
provides strong evidence in favor of a dynamic default rate.

To investigate the existence of dynamic covariates, we estimated the
model from Section \ref{sec3.1}. We once again assumed flat but proper priors as
$\gamma\sim U(0,1)$ and $\tau_{i} \sim G(0.001, 0.001)\ \forall i$. We
note here that even though the use of the flat Gamma prior for $\tau
_{i}$ is common in the literature, it has been known to have high
concentration at very small values and very low probability everywhere
else. However, our results were not influenced by the choice of the
prior. The MCMC estimation was less straightforward as opposed to the
model with static covariates. We ran the chain for 30,000 iterations as
the burn-in period and we collected 50,000 observations with a thinning
interval of 10. The mixing was slower since the full conditionals for
each time dependent regression coefficient require a Metropolis--Hasting
step. However, we did not encounter any convergence issues. In fact,
having dynamic regression coefficients improved both the fit and the
forecasting of the model as we discuss in the sequel.

\begin{table}[b]
\caption{Posterior statistics for $\bolds{\beta}$ and $\gamma$ of
the dynamic model with covariates} \label{summary1}
\begin{tabular*}{\textwidth}{@{\extracolsep{\fill}}lccccc@{}}
\hline
\textbf{Statistics} & $\bolds{\beta_{\mathrm{CMHPI}}}$ & $\bolds{\beta_{\mathrm{COFI}}}$ & $\bolds{\beta_{\mathrm{FOR}}}$ & $\bolds{\beta
_{\mathrm{Unemp}}}$ & $\bolds{\gamma}$ \\
\hline
25th & 0.0063 & 0.7003 & $-1.5430$ & 0.6252 & 0.2281 \\
Mean & 0.0160 & 0.8717 & $-1.3002$ & 0.8191 & 0.2466 \\
75th & 0.0256 & 1.0510 & $-1.0550$ & 1.0117 & 0.2643 \\
St. Dev & 0.0141 & 0.2663 & \phantom{$-$}0.3606 & 0.2826 & 0.0270 \\
\hline
\end{tabular*}
\end{table}

\begin{table}
\def\arraystretch{0.9}
\caption{$\operatorname{log} \{ p(N^{(t)}) \}$ and $\operatorname{log}(\mathrm{CPO})$ under
each model} \label{bftable}
 \begin{tabular*}{\textwidth}{@{\extracolsep{\fill}}lcccccc@{}}
\hline
& \textbf{DM1} & \textbf{DM2} & \textbf{DM3} & \textbf{DM4} & \textbf{DM5} & \textbf{BPM} \\
\hline
$\operatorname{log} \{ p(N^{(t)}) \}$ & $-579.99$ & $-577.61$ & $-577.67$ &
$-566.01$ & $-541.74$ & $-1416.28$ \\
$\operatorname{log}(\mathrm{CPO})$ & $-580.07$ & $-578.69$ & $-580.70$ & $-572.28$ & $-560.63$ & $-1372.62$ \\
\hline
\end{tabular*}  \vspace*{-3pt}
\end{table}

\begin{figure}[b]\vspace*{-3pt}

\includegraphics{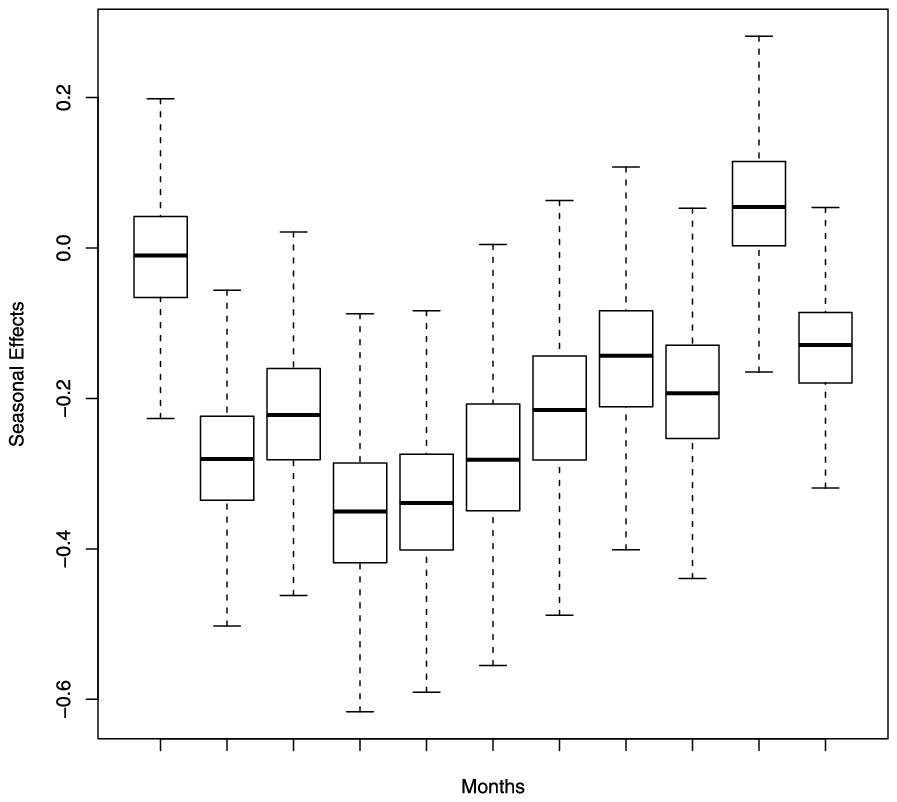}\vspace*{-3pt}

\caption{Boxplots for the seasonal effects from DM4.} \label{boxseas}
\end{figure}

\subsection{Model comparison}\label{sec5.4}
In order to compare the in-sample fit of the proposed models, we
computed the log-marginal likelihoods as in (\ref{bfmcmc2}) and the
conditional predictive ordinates in the log-scale as in\vadjust{\goodbreak} (\ref
{monteestimate}). In addition, we investigated presence of potential
polynomial trends by adding a second order polynomial to our dynamic
model with covariates and investigated the existence of seasonality by
including 11 dummy variables for each month with month 12 being the
reference period (as motivated by Figure \ref{defaultcount}). The
results are shown in Table \ref{bftable} where DM1 stands for the
dynamic model, DM2 for the dynamic model with covariates, DM3 for the
dynamic model with covariates and a second order polynomial trend, DM4
for the dynamic model with seasonality, DM5 for the dynamic model with
dynamic covariate effects and BPM for Bayesian Poisson regression
model. The dynamic model with dynamic covariate effects (DM5) has the
highest log-marginal likelihood value and the highest CPO with a Bayes
factor of $>$ 100 against its closest competitor ($\mathrm{BF}=\frac
{p(N^{(t)}|\operatorname{DM4})}{p(N^{(t)}|\operatorname{DM2})}$), which, according to
\citet{KR95}, shows decisive support in favor of DM5. The results
further support the lack of fit of the static model and show decisive
evidence in favor of the dynamic models with Bayes factors of $>$ 100
against the Poisson regression model. Furthermore, adding a second
order polynomial trend did not improve the model fit (the
log-likelihoods of DM2 vs. DM3 are identical). However, adding the
covariate information did improve the model fit (DM1 vs. DM2, DM3, DM4
and DM5). Even though the polynomial trend did not improve the model
fit for this particular data set, capturing the monthly periodic
effects did, as evidenced by the fit performance of model DM4 and the
boxplot of its seasonal effects of Figure \ref{boxseas}.\vadjust{\goodbreak}

In addition to understanding the default behavior of a given cohort, it
is also of interest to assess the model's ability to predict future
defaults for the cohort. In doing so, we considered two forecasting
horizons: one during the earlier stages of the cohort (between the 35th
and 44th months) and the second during the later stages (between the
135th and 144th months). For both forecast horizons, we sequentially
predicted the next month without using the information of the future.
For instance, we used the the first 34 months of data (both mortgage
counts and covariates when needed) as a training data set to predict
the 35th month and we sequentially predicted all 10 future months for
both forecast horizons.

To provide one-month-ahead forecasting comparisons, we mainly
considered two measures: the mean absolute percentage error (MAPE) and
the root mean squared error (RMSE) calculated as
%
\begin{equation}
\operatorname{MAPE} = \frac{1}{H} \sum_{t=1}^{H}
\frac{|N_{t} -
E(N_{t}|N^{(t-1)})|}{N_{t}},
\end{equation}
where $N_{t}$ is the actual default count observed during the month $t$
and $E(N_{t}|N^{(t-1)})$ is its one-month-ahead prediction. Similarly,
%
\begin{equation}
\operatorname{RMSE} = \sqrt{\frac{1}{H} \sum
_{i=1}^{H} \bigl\{ N_{t} - E
\bigl(N_{t}|N^{(t-1)}\bigr) \bigr\}^{2} },
\end{equation}
where $H$ represents the forecast horizon (in our example, $H=10$ for
both forecast intervals cases). We also considered other measures of
forecast performance such as the mean 95\% coverage probability and the
mean width of forecasts as in
\begin{eqnarray*}
 \operatorname{MCov}& =& \frac{1}{H} \sum_{t=1}^{H}
I\bigl\{ \bigl(N^{2.5}_{t}|N^{(t)}\bigr) <
N_{t} < \bigl(N^{97.5}_{t}|N^{(t)}\bigr)
\bigr\} \quad\mbox{and}
\\
 \operatorname{MWid}& =& \frac{1}{H} \sum_{t=1}^{H}
\bigl(N^{97.5}_{t}|N^{(t)}\bigr) -
\bigl(N^{2.5}_{t}|N^{(t)}\bigr),
\end{eqnarray*}
where $I(\cdot)$ is the indicator function, $(N^{97.5}_{t}|N^{(t)})$ and
$(N^{2.5}_{t}|N^{(t)})$ are 97.5th and 2.5th quantiles of the forecasts
for time $t$.

In addition to the dynamic models and the Poisson regression model, we
considered using a simple time series model such as the exponentially
weighted moving average (EWMA). To determine the smoothing constant for
EWMA, we minimized the mean absolute percentage deviations,
sequentially estimated it each time to predict the next month so that
the models were comparable.

\begin{table}
\tabcolsep=0pt
\caption{Forecasting performance comparison}\label{predictions}
\begin{tabular*}{\textwidth}{@{\extracolsep{\fill}}ld{3.2}d{2.2}d{2.2}d{2.2}d{2.2}d{2.2}d{2.2}d{2.2}d{2.2}d{2.2}d{3.2}d{2.2}@{}}
\hline
& \multicolumn{6}{c}{\textbf{Months 35--44}} & \multicolumn{6}{c@{}}{\textbf{Months
135--144}} \\[-6pt]
& \multicolumn{6}{c}{\hrulefill} & \multicolumn{6}{c@{}}{\hrulefill} \\
& \multicolumn{1}{c}{\textbf{DM1}} & \multicolumn{1}{c}{\textbf{DM2}} & \multicolumn{1}{c}{\textbf{DM4}} & \multicolumn{1}{c}{\textbf{DM5}} & \multicolumn{1}{c}{\textbf{BPM}} &
\multicolumn{1}{c}{\textbf{EWMA}} & \multicolumn{1}{c}{\textbf{DM1}} & \multicolumn{1}{c}{\textbf{DM2}} & \multicolumn{1}{c}{\textbf{DM4}} &
\multicolumn{1}{c}{\textbf{DM5}} & \multicolumn{1}{c}{\textbf{BPM}} &
\multicolumn{1}{c@{}}{\textbf{EWMA}}\\
\hline
MAPE & 13.6 & 17.09 & 13.4 & 17.8 & 22.1 & 21.2 & 36.9 & 57.5 & 58.2 &
51.5 & 103.8 & 65.0\\[3pt]
RMSE & 21.9 & 29.3 & 20.3 & 31.7 & 36.03 & 39.20 & 2.77 & 3.09 & 3.30 &
3.02 & 4.96 & 3.86\\[3pt]
MCov & 1.00 & 0.40 & 0.50 & 0.30 & 0.20 & \multicolumn{1}{c}{NA} & 1.00 & 0.10 & 0.20& 0.20
& 0.20 & \multicolumn{1}{c}{NA}\\[3pt]
MWid & 121 & 11.4 & 37.40 & 16.20 & 22.39 & \multicolumn{1}{c}{NA} & 13.4 & 2.15 & 1.70 &
2.36 & 4.29 & \multicolumn{1}{c}{NA}\\
\hline
\end{tabular*}
\end{table}

The results are shown in Table \ref{predictions} where DM1, DM2, DM4
and DM5 exhibit better forecasting performance than BPM and EWMA for
both forecast horizons. An interesting finding is that DM1 (model with
no covariates) provides better forecasts for one of these two
particular forecast horizons even though the overall model fit of DM2,
DM4 and DM5 were concluded to be superior in Table \ref{bftable}. In
addition, the mean coverage probabilities are both equal to 1 for DM1
since its prediction intervals are significantly wider than those of
DM2, DM4 and DM5. This might be due to the small values of $\gamma$
which imply high discounting, leading to high uncertainty in the
predictions. However, when we control for covariates and seasonal
factors such uncertainty is diminished as evidenced by the prediction
intervals given in Table \ref{predictions}. The narrowest prediction
intervals are provided by DM4 for the 35--44 horizon and by DM2 for the
135--144 horizon. Also, adding the seasonal components significantly
improved the forecasting performance of the dynamic models during the
35--44 month period where there is visual evidence of seasonality as
shown in Figure \ref{defaultcount}. Toward the end of the series during
the 135--144 month horizon, the default counts become more stable with
no obvious seasonal patterns. Thus, the dynamic model with no
covariates perform better due to its random walk type structure.

\section{Concluding remarks}\label{sec6}
In this paper we considered discrete time Bayes\-ian state space models
with Poisson measurements to model the aggregate mortgage default risk.
As pointed out by \citet{K11}, the Bayesian approach provides a coherent
framework to combine data with prior information and enables us to make
inferences using probabilistic reasoning. In addition, the proposed
state space models with stochastic default rate can capture the effects
of correlated defaults over time. In order to carry out the inference
of model parameters, we used Markov chain Monte Carlo methods such as
the Gibbs sampler, Metropolis--Hastings and forward filtering backward
sampling algorithms. In modeling the aggregate mortgage default risk,
we addressed whether the default rate was exhibiting static or dynamic
behavior and investigated the effects of macroeconomic variables on
default risk. Strong evidence in favor of dynamic default behavior at
the aggregate level was found. Furthermore, we found significant
effects of macroeconomic variables such as the regional conventional
mortgage home price index, federal cost of funds index, the homeowner
mortgage financial obligations ratio and the regional unemployment rate
on the aggregate mortgage default risk.\looseness=1

To the best of our knowledge, this is the first study using Bayesian
state space models considered in the mortgage default risk literature
at the aggregate level. Previous work mainly focuses on the individual
default behavior of borrowers, and the effects of mortgage loan,
property, borrower and economic characteristics on default risk. The
only study which considers modeling the mortgage default risk at the
aggregate level is due to \citet{T06}, who treats the default rate as a
stochastic process and points out the need for models that will assist
financial institutions to quantify mortgage default risk as required by
the Basel III framework. Kiefer (\citeyear{K10,K11}) introduces a Bayesian binomial
model for the default estimation of loan portfolios and their
estimation using expert information. Although neither work focuses on
mortgage default specifically, both highlight the need for models that
can capture correlated default behavior in a pool of loan portfolios
that are similar to mortgage cohorts consisting of several individual
borrowers. In addition, none of these studies consider the effects of
macroeconomic covariates on the default rate. In fact, \citet{T06}
comments on the lack and also on the need of covariate effects in his
model. Our proposed state space models can easily take into account
such covariate effects in both static and dynamic manners that are
crucial in volatile economic environments. Thus, the novelty of our
proposed models can be summarized as the introduction of Bayesian state
space models with Poisson measurements to the mortgage default risk
literature at the aggregate level, their ability to incorporate
correlated mortgage defaults and to capture the effects of covariates
on the default rate. In addition, the development of the Bayesian
Poisson state space models and their estimation using MCMC methods are
also modest contributions to the Poisson time series literature.

We believe that there are potential areas of research in modeling the
mortgage risk at the aggregate level that we would like to pursue in
the future. For instance, if information regarding the size of a given
were available, then one can consider correlated state space binomial
processes with both static and dynamic covariate effects to model the
default counts and compare them with those presented in this paper.
Although we did not encounter any efficiency issues in the use of MCMC
methods with our current data, another potential area is to consider
the use of particle filtering methods to speed up the convergence in
sequential updating and forecasting for larger data sets.\looseness=1


%



\printaddresses

\end{document}